\documentclass[12pt]{article}
\usepackage[colorlinks=true, linkcolor=blue, citecolor=blue, urlcolor=blue]{hyperref}
\usepackage{amsmath}
\usepackage{tikz}
\usetikzlibrary{arrows.meta, positioning, shapes.geometric, fit, calc}
\usepackage{float}
\usepackage{graphicx}
\usepackage[margin=1in]{geometry} % Better margins
\usepackage{changepage} % Add this to your preamble
\usepackage{pifont} % in the preamble
\usepackage{booktabs} % for \midrule, \bottomrule
\DeclareUnicodeCharacter{202F}{~} % maps U+202F to a normal non-breaking space

\newcommand{\cmark}{\ding{51}} % check mark
\newcommand{\xmark}{\ding{55}} % cross mark

\title{Filter-then-Verify: A Multiphase GNN and ModernBERT Framework for Social Engineering Detection in Email Networks}

\author{
	Barsat Khadka\thanks{Corresponding author. Email: \href{mailto:barsat.khadka@usm.edu}{barsat.khadka@usm.edu}}\textsuperscript{,a},
	Prasant Koirala\textsuperscript{a},
	Kshitiz Neupane\textsuperscript{a},
	Nick Rahimi\textsuperscript{a} \\[4pt]
	\textsuperscript{a}\,School of Computing Sciences and Computer Engineering, \\
	The University of Southern Mississippi, \\
	118 College Drive, Hattiesburg, MS 39406, USA
}

\begin{document}
	\date{}
	\maketitle

	%--------------------------------------------------
	% Highlights
	%--------------------------------------------------
	\section*{Highlights}
	\begin{itemize}
		\item Two-stage filter-then-verify pipeline combines inductive GNN with ModernBERT.
		\item Three-phase structural detection captures volume, relational, and contextual anomalies.
		\item Co-attention ModernBERT verifies long multi-message social engineering context.
		\item Specialized insider module uses linguistic drift to catch compromised accounts.
		\item Achieves 86\% structural recall and 92\% precision on augmented Enron dataset.
	\end{itemize}

	%--------------------------------------------------
	% Keywords
	%--------------------------------------------------
	\noindent\textbf{Keywords:} Phishing; GraphSAGE; ModernBERT; Co-attention; Anomaly~detection; Insider~threat; Email~security; Enron
	%--------------------------------------------------
	% Abstract
	%--------------------------------------------------
	\begin{abstract}
		Social engineering attacks exploit human trust rather than software vulnerabilities, making them difficult to detect using conventional filters. We propose a two-stage 'filter-then-verify' framework combining inductive Graph Neural Networks (GNNs) for structural anomaly detection with a co-attention ModernBERT model for content verification. The GNN identifies anomalous sender–receiver patterns, while BERT analyzes message context to reduce false positives. Using the Enron dataset augmented with realistic synthetic campaigns, we show that the framework achieves 86\% recall in structural filtering and over 92\% precision after BERT refinement, effectively detecting both external attacks and insider threats. Our results demonstrate that combining structural and content analysis allows practical, scalable detection of multi-stage social engineering attacks in email networks.
	\end{abstract}

	%--------------------------------------------------
	% Data Statement
	%--------------------------------------------------
	\section*{Data Availability Statement}
	The structural foundation of this study is the publicly available Enron Email Time-Series Network released by Miz et al.\ \cite{miz2018} on Zenodo (\href{https://doi.org/10.5281/zenodo.1342353}{https://doi.org/10.5281/zenodo.1342353}). The full source code for preprocessing, GraphSAGE training, three-phase structural anomaly scoring, the insider-threat module, and the co-attention ModernBERT verification stage can be released and will be made available from the corresponding author on reasonable request. However, the synthetic social engineering email corpus generated for this study, including the crafted attacker messages and follow-up communications used to train and evaluate the co-attention ModernBERT model, will not be released. These messages were carefully engineered to emulate realistic persuasion principles, authority cues, and multi-stage grooming tactics, and releasing them publicly carries a clear dual-use risk: they could be ingested into generative models or directly reused to craft convincing phishing and pretexting campaigns at scale. Withholding the attack corpus is therefore a deliberate safety measure consistent with responsible disclosure practices in offensive-security research.

	%--------------------------------------------------
	% Introduction
	%--------------------------------------------------
	\section{Introduction}
	Social engineering attacks, such as phishing and pretexting, are now the most common type of cyber attacks in corporate environments. Unlike traditional attacks that exploit software vulnerabilities, these attacks target people first, gaining access to systems before exploiting software. For example, attackers may trick employees into granting access by slowly building trust, often by pretending to be a boss, IT support, or another trusted figure. Since these messages often look normal and contain no malware, traditional email filters based only on content often fail to detect them.
	A major challenge in detecting social engineering attacks is that malicious behavior often emerges before the message content itself becomes obviously suspicious. Early warning signs are instead reflected in communication patterns, such as unusual sender–receiver relationships, abnormal email frequency, or interactions involving previously unseen external accounts. These structural signals are difficult to capture using traditional natural language processing approaches alone.
	
	In this work, we propose an inductive Graph Neural Network (GNN) framework to detect social engineering attacks by modeling email communication as a graph and identifying anomalous patterns in who communicates with whom, when, and how frequently. This first stage introduces a three-phase structural anomaly detection pipeline. Potentially malicious interactions are identified using: (1) volume-based spike detection, (2) relational anomaly analysis that contrasts historical interaction frequency with embedding-based structural similarity, and (3) community-level contextual analysis. To reduce false positives and confirm genuine attacks, we then apply a BERT-based content analysis stage that verifies the semantic intent of the flagged communications.
	
	Prior research on social engineering detection remains limited and fragmented. Some approaches rely on predefined security dictionaries or handcrafted intent scores, while others apply traditional machine learning techniques based on shallow linguistic features.\cite{seader} These methods often struggle to generalize to targeted or multi-stage attacks, where adversaries adapt their language over time. Additionally, many graph-based anomaly detection techniques are transductive, making them unsuitable for real-world email environments where new users and external attackers continuously appear.
	
	The primary contribution of this paper is a practical two-stage filter-then-verify architecture that balances detection accuracy with operational feasibility. The inductive GNN component provides high recall in dynamic email networks by identifying structurally anomalous interactions, while the BERT-based verification stage significantly improves precision by reducing false positives through contextual analysis. Using the Enron email dataset augmented with realistic synthetic social engineering campaigns, we demonstrate that the proposed framework effectively detects a wide range of attack scenarios, including both external attackers and compromised internal accounts.

	%--------------------------------------------------
	% Literature Review
	%--------------------------------------------------
	\section{Related Work}
	Graph Neural Networks (GNNs) and BERT have been extensively studied across a wide range of tasks. However, their application in detecting social engineering attacks remains relatively unexplored. This section reviews existing approaches that model anomaly detection as a graph-based problem, as well as content-based anomaly detection methods, to discuss their strengths and limitations from the viewpoint of detecting social engineering attacks in an email network.
	\subsection{Graph-based Anomaly Detection}
	\subsubsection{Non-neural and Incremental Methods}
	Early studies on the Enron email corpus used classical scan statistics over fixed-radius neighborhoods to identify unusual activity patterns.\cite{priebe2005} Later work such as AnomRank introduced fast online algorithms that monitor changes in node importance scores to identify structural and weight-based anomalies in dynamic graphs.\cite{anomrank} While these approaches are computationally efficient, they rely only on graph structure. They do not use rich node features and cannot handle entirely new external nodes, which limits their usefulness for detecting social engineering attacks.
	
	\subsubsection{Transductive Graph methods for anomaly detection}
	Ding et al. (2019) proposed a framework DOMINANT, a deep anomaly detection framework using a graph convolutional autoencoder that uses a GCN encoder to create low-dimensional embeddings using both structure and attributes where anomalies are nodes with high reconstruction error. \cite{ding2019} DOMINANT's strengths include combining structure and attributes in a deep model, beating simpler methods like ego-network or subspace analysis on datasets like social media or gene networks. However, being transductive limits it for real-time email systems where new attackers might appear after training. It also uses reconstruction errors without specific community or time-based checks, which might cause more false alarms in email timelines.
	\subsubsection{Inductive Graph Representation Learning for Anomaly Detection}
	Inductive graph representation learning has gained significant attention for anomaly and fraud detection due to its ability to generate embeddings for previously unseen nodes using existing training data, without requiring retraining on the entire graph. This capability is especially important in social engineering as new senders frequently appear even after model has been deployed.
	
	A relevant work is that of Van Belle et al. (2022), who applied GraphSAGE and Fast Inductive Graph Representation Learning (FI-GRL) to real-world credit card transaction networks.\cite{van2022}  The study showed that when traditional transaction attributes are provided as node features during GraphSAGE training, classification performance increases even under extreme class imbalance. However, their approach is tailored to bipartite (or tripartite) transaction networks and treats fraud as node classification on repeated transactions. In contrast, email communication graphs are non-bipartite, and social engineering attacks usually appear as rare, one-way interactions from previously unseen external senders rather than a repeated occurrence.
	\subsection{Content based Social Engineering Attack Detection}
	Previous works on content-based detection for social engineering use shallow machine learning models utilizing linguistic features \cite{alsufyani2021}. While these methods work well against mass-phishing, they fail to capture the relationship and context between the two parties involved in a social engineering attempt. Ilias et al. (2022) \cite{ilias2022} demonstrated that adding co-attention to BERT-based models helps them capture the connection between an attacker’s first message and later follow-up messages, leading to better detection accuracy.
	
Building on this foundation, we employ ModernBERT \cite{warner2024} with a co-attention mechanism, originally proposed by Lu et al. \cite{lu2016hierarchical} for visual question answering and later adapted for deception detection by Ilias et al. \cite{ilias2022}. ModernBERT offers several practical improvements over the original BERT model used by Ilias et al. \cite{ilias2022}, most notably support for much longer input sequences (up to 8,192 tokens versus BERT’s 512). This is especially important for social engineering detection, where attacks often unfold over multiple messages and require understanding long conversational context rather than a single email. To handle this message history efficiently, we also use a lightweight summarization model to condense past communications into a concise context. This summary is then provided alongside the current message to the co-attention architecture.
	\subsection{Gap and Our Study Motivation}
	Existing graph-based methods, like DOMINANT, struggle with new attackers because they require retraining on the entire graph. This is a major limitation for social engineering detection, where attackers often use previously unseen accounts. While GraphSAGE has been effective for fraud detection in credit card networks, it is not tailored for email networks, which are non-bipartite and highly temporal. On the content side, traditional NLP approaches rely on shallow linguistic features and fail to capture the evolving context across multiple messages. This highlights the need for a framework that integrates inductive structural learning with  context-aware content analysis to effectively detect multi-stage social engineering attacks.
    
    A significant challenge in social engineering research is the scarcity of public enterprise email datasets. A common concern when using historical datasets such as Enron (1999–2002) is whether they generalize to modern social engineering attacks. We argue that unlike other cyber-exploits, social engineering attacks target fundamental human trust patterns and organizational hierarchies that are structurally time-invariant to a large extent.
    
    While message semantics play a critical role in our verification phase, they are utilized only after the initial structural filtering. Our experiments demonstrate that using pure graph structure , which represents the network strictly as nodes and edges without message content or metadata of the sender-receiver pair, it yielded a recall of 86\%. This high detection rate using only topological features indicates that the primary signals of social engineering are embedded in communication structure. Since these structural patterns of trust and anomaly remain stable over time, the age of the dataset does not compromise the validity of the structural detection evaluation. To validate this, we performed a temporal split—training on 1999–2000 data and testing on 2001–2002. The model’s consistent performance confirms that these structural anomalies remain stable across time.

	%--------------------------------------------------
	% Methodology
	%--------------------------------------------------
	\section{Experimental Setup}
	This section describes the dataset used for evaluation, the construction of the email communication graph, the process of synthetic attack injection, and the architectural details of the proposed GraphSAGE and BERT-based components. We also detail the three-phase structural anomaly detection pipeline and the final score aggregation strategy.
	
	\subsection{Dataset and Graph Construction}
	
	We use the preprocessed Enron Email Time-Series Network released by Miz et al. \cite{miz2018} as the foundation for our experiments. The dataset provides email addresses as nodes, sender–receiver relationships as directed edges, and, for each node, a record of the days emails were sent and the corresponding counts. Since this format is not directly suitable for our model, we convert each record into a fixed-length vector of $T = 1{,}448$ days, where each entry $t \in [1, T]$ represents the number of emails sent on that day (zero if none were sent). Formally, for a node $v \in V$, this gives a feature vector $\mathbf{x}_v \in \mathbf{R}^{T}$.
	
	These vectors serve as node features, and together with the directed edge set $E \subseteq V \times V$, they form the attributed communication graph $G = (V, E, \{\mathbf{x}_v\}_{v \in V})$. The graph contains approximately $|V| = 6{,}599$ nodes and captures email activity from January 1, 1999, to July 30, 2002. This setup lets our GraphSAGE model use both the email activity over time and the connections between users.
	
	\subsection{Synthetic Attack Injection}
	The original Enron dataset contains no ground-truth labels for social engineering attacks. So we construct 5 sets of realistic campaigns based on social engineering principles \cite{Mouton} and some examples \cite{Qualcomm} \cite{Zelster} and inject them into the current enron graph. The GraphSAGE model is trained only on the original clean Enron graph. After training, the synthetic attackers (new nodes) and their emails/replies are injected as directed edges. This setup utilizes GraphSAGE's inductive ability, which learns a general aggregation function to compute embeddings\cite{hamilton2017inductive}.This prevents retraining whenever a new node is added to the graph. 
	\begin{table}[h!]
		\centering
		\caption{Synthetic Attack Campaigns}
		\small
		\begin{tabular}{|c|c|c|c|c|}
			\hline
			\textbf{Campaign} & \textbf{Attacker} & \textbf{Target(s)} &
			\textbf{Duration} & \textbf{Emails} \\ \hline
			
			C1 & 6600 & 97 & Jul 20--31, 1999 & 11 \\ \hline
			C2 & 6601 & 171 & Aug 4--15, 1999 & 11 \\ \hline
			C3 & 6602 & 221, 223, 225, 226, 227 & Oct 28--Nov 2, 1999 & 18 \\ \hline
			C4 & 6603 & 817, 1017 & Feb 20, 1999--Feb 15, 2000 & 13 \\ \hline
			C5 & 1023 & 145 & Apr 3--Sep 18, 2000 & 24 \\ \hline

		\end{tabular}
		\label{tab:synthetic_campaigns}
	\end{table}
	
	Authors in \cite{Mouton} \cite{Khadka} identify specific compliance principles like authority, urgency , trust as to why a target complies with a request. C1 is based on Friendship or Liking principle. C2 was inspired by \cite{Qualcomm} where an attacker sent malicious invoice by email and quickly followed up with a phone call posing as a senior executive, using authority and urgency. We simulated the follow-up call by adding a burst of  one-way emails right after the initial one. C3 was inspired by \cite{Zelster}, where attackers placed notices on parked vehicles directing many victims to a malicious website. We modeled this attack as a broadcasting pattern by a single external node that sends one‑way emails to multiple unrelated users. C4 represents a low and slow attack based on principles of consistency\cite{Khadka}. C5 represents an insider or compromised account scenario which is the most difficult to detect.
	\subsection{Architecture}

	% FIGURE 1: Training Phase (Original Data) - Using [H] to force position
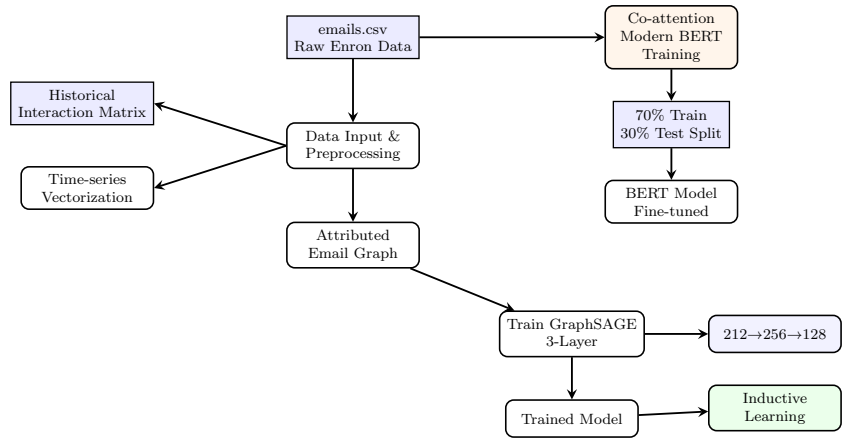
\begin{figure}[H]
	
	\begin{adjustwidth}{0cm}{0cm}
		\centering
		\begin{tikzpicture}[
			scale=0.70,
			transform shape,
			node distance=0.6cm and 1.2cm,
			every node/.style={font=\scriptsize},
			filebox/.style={
				draw,
				rectangle,
				minimum width=2.0cm,
				minimum height=0.7cm,
				fill=blue!8,
				align=center,
				line width=0.5pt
			},
			process/.style={
				draw,
				rectangle,
				rounded corners=3pt,
				align=center,
				minimum width=2.5cm,
				minimum height=0.7cm,
				fill=white,
				line width=0.6pt
			},
			arrow/.style={->, >=stealth, line width=0.7pt}
			]
			
			% ORIGINAL EMAILS CSV
			\node[filebox] (emails_csv) {emails.csv\\Raw Enron Data};
			
			% Preprocessing - centered
			\node[process, below=1.2cm of emails_csv] (preprocess) {Data Input \&\\Preprocessing};
			
			% Time-series - LEFT side, lower
			\node[process, left=2.5cm of preprocess, yshift=-0.8cm] (ts_vector) {Time-series\\Vectorization};
			
			% Matrix - LEFT side, upper
			\node[filebox, left=2.5cm of preprocess, yshift=0.8cm] (matrix) {Historical\\Interaction Matrix};
			
			% Graph building - below preprocessing (SHORTER DISTANCE)
			\node[process, below=1.0cm of preprocess] (graph_build) {Attributed\\Email Graph};
			
			% GraphSAGE Training
			\node[process, below right=0.8cm and 1.5cm of graph_build] (sage_train) {Train GraphSAGE\\3-Layer};
			\node[process, below=0.8cm of sage_train] (sage_model) {Trained Model};
			
			% Model details
			\node[process, right=1.2cm of sage_train, fill=blue!5] (details) {212→256→128};
			\node[process, below=of details, fill=green!8] (inductive) {Inductive\\Learning};
			
			% CO-ATTENTION BERT TRAINING
			\node[process, right=3.5cm of emails_csv, fill=orange!8] (coattention) {Co-attention\\Modern BERT\\Training};
			\node[filebox, below=0.6cm of coattention] (train_test) {70\% Train\\30\% Test Split};
			\node[process, below=0.6cm of train_test] (bert_model) {BERT Model\\Fine-tuned};
			
			% ARROWS
			% Main pipeline arrows
			\draw[arrow] (emails_csv) -- (preprocess);
			
			% Split arrows from preprocessing - both to left side
			\draw[arrow] (preprocess.west) -- (ts_vector.east);
			\draw[arrow] (preprocess.west) -- (matrix.east);
			
			% Down to graph (SHORTER)
			\draw[arrow] (preprocess) -- (graph_build);
			
			% Training flow
			\draw[arrow] (graph_build) -- (sage_train);
			\draw[arrow] (sage_train) -- (sage_model);
			\draw[arrow] (sage_train) -- (details);
			\draw[arrow] (sage_model) -- (inductive);
			
			% BERT training flow
			\draw[arrow] (emails_csv.east) -- ++(1.0,0) |- (coattention.west);
			\draw[arrow] (coattention) -- (train_test);
			\draw[arrow] (train_test) -- (bert_model);

		\end{tikzpicture}
	\end{adjustwidth}
	\caption{Training Phase: Raw emails.csv is preprocessed into time-series features and graph structure. GraphSAGE is trained for structural analysis while co-attention BERT is trained on email content. Both models support inductive learning.}
	\label{fig:training_phase}
\end{figure}

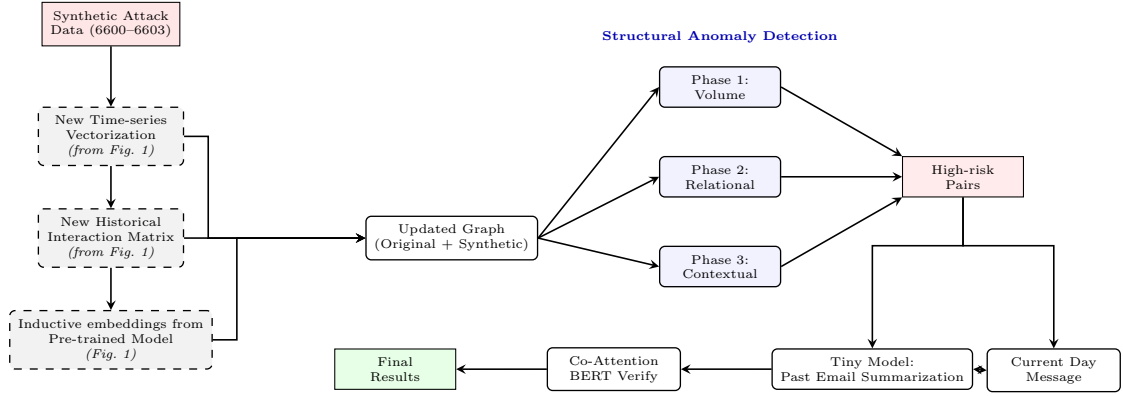
\begin{figure}[H]
	\begin{adjustwidth}{0cm}{0cm}
		\centering
		\begin{tikzpicture}[
			scale=0.8,
			transform shape,
			node distance=0.9cm and 1.5cm,
			every node/.style={font=\tiny},
			filebox/.style={
				draw,
				rectangle,
				minimum width=2.0cm,
				minimum height=0.55cm,
				fill=red!8,
				align=center,
				line width=0.4pt
			},
			process/.style={
				draw,
				rectangle,
				rounded corners=2pt,
				align=center,
				minimum width=2.2cm,
				minimum height=0.65cm,
				fill=white,
				line width=0.5pt
			},
			phasebox/.style={
				process,
				fill=blue!5,
				minimum width=2.0cm
			},
			refbox/.style={
				process,
				dashed,
				fill=gray!10,
				minimum width=2.4cm
			},
			arrow/.style={->, >=stealth, line width=0.6pt},
			interact/.style={<->, >=stealth, dashed, line width=0.5pt}
			]
			% INPUT
			\node[filebox, fill=red!10] (synthetic_data)
			{Synthetic Attack\\Data (6600--6603)};
			% VERTICAL STACK
			\node[refbox, below=1.0cm of synthetic_data] (synthetic_ts)
			{New Time-series\\Vectorization\\\textit{(from Fig.~1)}};
			\node[refbox, below=0.7cm of synthetic_ts] (historical_matrix)
			{New Historical\\Interaction Matrix\\\textit{(from Fig.~1)}};
			\node[refbox, below=0.7cm of historical_matrix] (pretrained)
			{Inductive embeddings from \\Pre-trained Model\\\textit{(Fig.~1)}};
			% UPDATED GRAPH
			\node[process, right=3.0cm of historical_matrix] (updated_graph)
			{Updated Graph\\(Original + Synthetic)};
			% ARROWS FROM INPUT TO GRAPH
			\draw[arrow] (synthetic_data.south) -- (synthetic_ts.north);
			\draw[arrow] (synthetic_ts.south) -- (historical_matrix.north);
			\draw[arrow] (historical_matrix.south) -- (pretrained.north);
			\draw[arrow] (synthetic_ts.east) -- ++(0.4,0) |- (updated_graph.west);
			\draw[arrow] (historical_matrix.east) -- ++(0.4,0) |- (updated_graph.west);
			\draw[arrow] (pretrained.east) -- ++(0.4,0) |- (updated_graph.west);
			% PHASES
			\node[phasebox, right=2.0cm of updated_graph, yshift=2.5cm] (p1)
			{Phase 1:\\Volume};
			\node[phasebox, below=0.8cm of p1] (p2)
			{Phase 2:\\Relational};
			\node[phasebox, below=0.8cm of p2] (p3)
			{Phase 3:\\Contextual};
			% STRUCTURAL LABEL
			\node[above=0.25cm of p1, font=\tiny\bfseries, text=blue!70!black]
			{Structural Anomaly Detection};
			% GRAPH TO PHASES
			\draw[arrow] (updated_graph.east) -- (p1.west);
			\draw[arrow] (updated_graph.east) -- (p2.west);
			\draw[arrow] (updated_graph.east) -- (p3.west);
			% HIGH-RISK PAIRS (aligned horizontally with phases)
			\node[filebox, right=2.0cm of p2] (pairs)
			{High-risk\\Pairs};
			\draw[arrow] (p1.east) -- (pairs.north west);
			\draw[arrow] (p2.east) -- (pairs.west);
			\draw[arrow] (p3.east) -- (pairs.south west);
			% BOTTOM ROW - all on same level
			% TINY MODEL SUMMARIZATION (positioned below and left of pairs)
			\node[process, below=2.5cm of pairs, xshift=-1.5cm] (summary)
			{Tiny Model:\\Past Email Summarization};
			% CURRENT DAY MESSAGE (positioned below and right of pairs)
			\node[process, below=2.5cm of pairs, xshift=1.5cm] (current_msg)
			{Current Day\\Message};
			% CO-ATTENTION BERT (left of tiny model)
			\node[process, left=1.5cm of summary] (bert)
			{Co-Attention\\BERT Verify};
			% FINAL RESULTS (leftmost)
			\node[filebox, left=1.5cm of bert, fill=green!10] (final)
			{Final\\Results};
			% Arrows from High-risk Pairs - going left and right
			\draw[arrow] (pairs.south) -- ++(0,-0.8) -| (summary.north);
			\draw[arrow] (pairs.south) -- ++(0,-0.8) -| (current_msg.north);
			% Horizontal interaction between Tiny Model and Current Message
			\draw[interact] (summary.east) -- (current_msg.west);
			% Arrow from Tiny Model to BERT
			\draw[arrow] (summary.west) -- (bert.east);
			% Arrow from BERT to Final Results
			\draw[arrow] (bert.west) -- (final.east);
		\end{tikzpicture}
	\end{adjustwidth}
	\caption{Detection phase with synthetic attacks. Structural anomaly detection is performed in three phases to identify high-risk pairs. Past emails are summarized using a Tiny Model (T5-small \cite{raffel2020}) and combined with the current day's message. Both interact horizontally, and the Tiny Model output is fed into a Co-Attention BERT model to produce the final detection results.}
	\label{fig:detection_phase}
\end{figure}
	\subsection{Three-Phase Structural Anomaly Detection}
Our framework adopts a three-phase approach to detect structural anomalies in an email communication graph. Each phase targets a distinct aspect of potential social engineering behavior, and the resulting scores are normalized and aggregated to produce a unified anomaly score. Specifically, the three phases capture 1) deviations in individual email activity 2) abnormal sender–receiver relationships, 3) contextual deviations based on how a user’s activity compares to the overall level of activity in their community on that day.
	\subsubsection{Phase 1: Spike Detection (Email Volume Anomaly)}
	In the first phase, we detect abnormal email volume for each sender by comparing current activity against historical behavior. For a sender node $v$, let $\{x_v^1, x_v^2, \dots, x_v^{t-1}\}$ denote the historical daily email counts, and $x_v^t$ denote the count on the current day $t$. The mean $\mu_v$ and standard deviation $\sigma_v$ are computed using the historical values up to day $t-1$, and the z-score for day $t$ is calculated as
	\[z_v^t = \frac{x_v^t - \mu_v}{\sigma_v}.
	\] This score measures how much the sender’s current activity deviates from their usual behavior. Since social engineering attacks often involve bursts of outreach to different recipents to establish contact and find a potential target, unusually high z-scores are indicative of suspicious behavior. The z-score is passed through a Gaussian Error Function to obtain a normalized volume anomaly score $S_1(v) \in [0,1]$.
\subsubsection{Phase 2: Neighborhood and Interaction Analysis (Relational Anomaly)}

Phase 2 detects anomalous relationship formation patterns by combining embedding similarity with historical interaction pattern into a contrast aware relational score. This allows us to identify interactions that are structurally unusual or inconsistent with past communication behavior. 

	\paragraph{Historical Interaction Score:}  
	For each sender–recipient pair on a given day, we measure the strength of their past relationship. This is done by counting the number of days on which at least one email was exchanged between the pair compared to the number of days without interaction. We also introduce a dynamic threshold K for normalization. K is the median number of past email interactions across all pairs. Formally, the raw historical interaction score \(h_{vu}\) is defined as
	
	\[
	h_{vu} =
	\begin{cases}
		\displaystyle \frac{\text{\# days $v$ emailed $u$ in the past}}{\text{\# days $v$ did not email $u$ in the past}}, & \text{if at least one past interaction exists,} \\[1mm]
		0, & \text{if $v$ has never emailed $u$ before (new interaction).}
	\end{cases}
	\]
	
	Since \(h_{vu}\) can take unbounded values, we normalize it to a bounded frequency
	
	\[
	f_{vu} = \frac{h_{vu}}{K_{\text{dynamic}} + h_{vu}} \in [0,1]
	\]

\paragraph{Embedding-Based Structural Similarity:}  
Next, we consider how this interaction of the sender-receiver fits within the overall communication graph. We compute the cosine similarity between the embeddings of the sender and recipient:

\[
\text{sim}_{vu} = \frac{\mathbf{z}_v \cdot \mathbf{z}_u}{\|\mathbf{z}_v\| \|\mathbf{z}_u\|} \in [0,1].
\]

A high similarity means the sender and recipient behave similarly in the network according to the 3-layer GraphSAGE embeddings, while a low similarity indicates an unusual interaction.

\paragraph{Contrast-Aware Relational Score}  
While historical interaction frequency and embedding-based structural similarity provide useful insights individually, neither alone is sufficient to reliably detect unusual relationships. Being close in the network does not necessarily mean that two nodes have interacted frequently, and frequent interactions may still behave unexpectedly. To address this, we combine both measures into a contrast-aware relational score.

\[
S_2(v \to u) = \alpha \cdot (1 - f_{vu}) + \beta \cdot (1 - \text{sim}_{vu}) + \gamma \cdot f_{vu} \cdot (1 - \text{sim}_{vu}),
\]

where \(\alpha, \beta, \gamma\) control the relative contributions of each term. The table below illustrates the contribution of each different combinations.

\begin{table}[h!]
	\centering
	\small
	\begin{tabular}{|l|c|c|l|c|c|}
		\hline
		\textbf{Case} & \(f_{vu}\) & \(\text{sim}_{vu}\) & \textbf{Term contributions} & \textbf{S2 Score} & \textbf{Suspicious?} \\ \hline
		New / rare        & low    & any      & high from $\alpha(1-f)$ & high     & \cmark \\ \hline
		Frequent, usual   & high   & high     & all low                  & low      & \xmark \\ \hline
		Frequent, unusual & high   & low      & contrast term $\gamma f (1-\text{sim})$ high & high & \cmark \\ \hline
		Moderate balanced & medium & high     & moderate                 & low      & \xmark \\ \hline
	\end{tabular}
	\caption{Summary of relational anomaly detection based on historical frequency and embedding similarity.}
	\label{tab:phase2_cases}
\end{table}

	\subsubsection{Phase 3: Community Detection (Contextual Anomaly)}
In Phase 3, we detect contextual anomalies by analyzing sender's engagement with the community of the different recipients on the given day. We first convert the email graph into an undirected graph and apply the Louvain algorithm to partition nodes into communities. The Louvain algorithm is a widely used community detection method that identifies groups of nodes by maximizing modularity in a greedy manner \cite{blondel2008}. For each sender–recipient pair, we measure how frequently the sender interacts with the recipient’s community relative to how frequently it interacts with other communities on the recipient list.

\[
\text{CommFrac}_{v \to u} =
\frac{\# \text{ recipients of } v \text{ in the community of } u}
{\# \text{ total recipients of } v}.
\]

A lower value indicates that the sender rarely interacts with the recipient’s community.We then compute the Phase 3 anomaly score $S_3$ for each sender--recipient pair using linear inversion:
\[
S_3 = 1 - \text{CommFrac}_{v \to u}.
\]
 While the Louvain algorithm provides an unsupervised way to identify communities from graph structure alone, we hypothesize that using predefined labeled communities may further improve the effectiveness of this phase. We leave this exploration to future work.

\subsection{Specialzed detection for insider threats}
A unique challenge arises when detecting social engineering attacks initiated by compromised internal accounts (e.g., Campaign 5). Unlike external attackers, who usually appear as new nodes with no history, compromised insiders have high centrality, established trust relationships, and a history of high-volume communication. To address this, we introduce a conditional branching logic. If an interaction involves two established internal nodes, we shift from global anomaly detection to an ego-centric baseline comparison. This specialized detection modifies the structural phases as follows:

\subsubsection{Modified Phase 1: } Standard volume detection analyzes the sender's total daily output. For insiders, this is ineffective because a single malicious email is hidden by legitimate traffic. Instead, we modify Phase 1 to calculate the z-score of the pairwise interaction volume between the specific sender and recipient. This focuses the anomaly detection on the relationship rather than the user.
\subsubsection{Substitution of Phase 2: Linguistic Drift Analysis}
For insiders, the Phase 2 structural scores (embedding similarity) are deceptively low risk because the compromised node is structurally legitimate. To compensate, we substitute the relational structural score with a \textit{Linguistic Drift} metric ($D_{ling}$). We compute the cosine distance between the ModernBERT embedding of the current message and the centroid of the user's last 50 sent messages. This detects the semantic style shift that occurs when an external adversary assumes control of a legitimate account. For the Linguistic Drift metric, we utilize frozen embeddings from pre-trained ModernBERT. Fine-tuning was reserved for the final Classification Stage to prevent overfitting the anomaly detector to specific keywords.

\subsubsection{Insider Risk Aggregation}
The final Insider Risk Score is computed as a weighted sum of these targeted components, prioritizing deviations from the user's personal history:
\begin{equation}
    S_{insider} = 0.3 \cdot D_{rec} + 0.4 \cdot D_{ling} + 0.2 \cdot I_{man} + 0.1 \cdot S_{struct}
\end{equation}
where $D_{rec}$ represents the Pairwise Recipient Deviation (calculated in Modified Phase 1). $D_{ling}$ is the Linguistic Drift, measured using the frozen ModernBERT embeddings prior to classification. $I_{man}$ represents the specific Manipulation Intent score ($0$ to $1$) outputted by the fine-tuned BERT classifier, and $S_{struct}$ is the baseline cosine similarity between the sender and recipient nodes derived from the GNN. 

	\subsection{Content-Based Verification Using Co-Attention BERT}
		\subsubsection{ModernBERT architecture}
	After the structural anomaly detection pipeline identifies high-risk sender-receiver pairs, we apply a content-based verification layer to reduce false positives and confirm genuine social engineering attempts. This section details the architecture and mathematical formulations of our ModernBERT + Co-Attention model. 
	
For each email in a sender--receiver pair, we prepare two input sequences: $x_1$, which is the summarized history of all past messages (generated by the Tiny Model Summarizer), and $x_2$, which contains the concatenated subject and body text of the current email. These sequences are tokenized and passed through the ModernBERT model to obtain contextualized embeddings.
Mathematically, 
\[
H_1 = \text{ModernBERT}(x_1) \in \mathbf{R}^{N \times d}, \quad
H_2 = \text{ModernBERT}(x_2) \in \mathbf{R}^{T \times d},
\]
where $N$ and $T$ are the sequence lengths, and $d = 768$ is the hidden dimension of ModernBERT-base.
\subsubsection{Co-Attention Mechanism and Classification}
Following Ilias et al. \cite{ilias2022}, we employ a co-attention mechanism to capture semantic dependencies between the current message and the summarized history.  

We first compute an affinity matrix:
\[
F = \tanh(H_1^\top W_l H_2),
\]
where $W_l$ is a learned weight matrix. This matrix is used to derive attention maps:
\[
H_s = \tanh(W_s H_2 + (W_c H_1) F), \quad
H_c = \tanh(W_c H_1 + (W_s H_2) F^\top).
\]

The attention maps are converted into probability distributions $a_s$ and $a_c$ via softmax, which are then used to obtain co-attended vectors $\hat{s}$ and $\hat{c}$. The final representation is
\[
z = [\hat{s}, \hat{c}].
\]

This vector is passed through a four-layer MLP with GELU activations, layer normalization, and 0.3 dropout. The architecture follows a decreasing bottleneck:
\[
1536 \to 512 \to 256 \to 128 \to 1,
\]
with the final probability computed as 
\[
p = \sigma(W_4 h_3 + b_4).
\]

\subsubsection{Training Strategy}
We employ a two-stage training strategy to optimize the model:

\begin{enumerate}
	\item \textbf{Frozen ModernBERT Phase:} Initially, we freeze the ModernBERT parameters and train only the co-attention mechanism and classification layers. This allows the task-specific components to adapt to the pre-trained representations without catastrophic forgetting.
	
	\item \textbf{Fine-Tuning Phase:} After the co-attention and classification layers converge, we unfreeze the ModernBERT parameters and perform end-to-end fine-tuning with a lower learning rate. This allows the entire model to jointly optimize for social engineering detection.
\end{enumerate}

The model is trained using binary cross-entropy loss with class weights to handle the imbalanced nature of the dataset, where legitimate emails significantly outnumber social engineering attempts.

	\subsection{Score Aggregation}
A tested limitation of independently summing the phase scores is that it produces a large number of false positives. One such case is the flagging of burst announcement emails sent by administrators. In corporate environments like enron, administrators frequently send high-volume emails to a wide set of recipients, across multiple departments (communities). If an administrator has historically interacted with these recipients over long periods, their sender–receiver relationships are structurally normal and should not be treated as suspicious. However, an independent linear aggregation always allows high volume (phase 1) to dominate the final anomaly score.

To address this, we use the volume as a conditional multiplier on a base risk score. The base risk is derived from combining the scores of phase 2 and phase 3, with appropriate weights. This way, high volume only increases the anomaly score when the underlying interactions are actually unusual. Mathematically, 
\[
S_{\text{final}} = (1+ 
w_1 \cdot S_1) \cdot \big( w_2 \cdot S_2 + w_3 \cdot S_3 \big).
\]

	%--------------------------------------------------
	% Results
	%--------------------------------------------------
\section{Results}

The proposed framework outputs an aggregated anomaly score $S_{\text{final}} \in [0,1]$ for each sender--receiver interaction. To label an interaction as anomalous, a threshold score must be defined. During experiments, campaign-related sender--receiver interactions consistently produced scores in the range of $0.65$ to $0.95$. One exception occurred in Campaign~1, where an attack interaction scored $0.63$.

Lowering the global threshold to capture this single case resulted in a significant increase in false positives and was therefore impractical. To analyze the trade-off between threshold score and false-positives, we evaluate structural framework performance at four thresholds scores: $0.65$, $0.70$, $0.75$, and $0.80$.

\subsection{Campaign wise Structural Detection}

\begin{table}[h!]
	\centering
	\small
	\begin{tabular}{|l|c|c|c|c|c|}
		\hline
		\textbf{Campaign} & \textbf{Total} & \textbf{$\ge$0.65} & \textbf{$\ge$0.70} & \textbf{$\ge$0.75} & \textbf{$\ge$0.80} \\ \hline
		C1 & 6  & 5 (83.3\%)  & 5 (83.3\%)  & 2 (33\%)  & 1 (17\%) \\ \hline
		C2 & 9  & 9 (100\%)   & 9 (100\%)   & 9 (100\%) & 9 (100\%) \\ \hline
		C3 & 10 & 10 (100\%)  & 9 (90\%)    & 5 (50\%)  & 5 (50\%) \\ \hline
		C4 & 8  & 8 (100\%)   & 8 (100\%)   & 7 (87.5\%)& 7 (87.5\%) \\ \hline
		C5 & 24 & 21 (87.5\%) & 18 (75.0\%) & 10 (41.7\%) & 6 (25.0\%) \\ \hline
		\textbf{Total} & \textbf{57} & \textbf{53 (93.0\%)} & \textbf{49 (86.0\%)} & \textbf{33 (57.9\%)} & \textbf{28 (49.1\%)} \\ \hline
	\end{tabular}
	\caption{Campaign-wise Structural Detection Performance Across Thresholds}
	\label{tab:detection_performance}
\end{table}

The GNN model achieves a good recall score. However, achieving higher recall comes at the expense of lower precision, as shown in Table \ref{tab:gnn_metrics}. In the context of our system, the main goal of the GNN is to maximize recall, which it does effectively, even if precision is weak. The Co-Attention BERT model addresses the issue of precision. As shown in Table \ref{tab:bert_metrics}, BERT significantly improves precision across campaigns, with campaign 3 slightly lower at 95\%. 

Campaign 5 represents a compromised account or insider threat, where the attacker was a legitimate user present in the original training data. Because the historical interactions for this node were structurally and linguistically normal, the BERT model faced significant challenges in distinguishing the synthetic attack messages from the established good baseline. This leads to lower detection performance compared to external campaigns, as the co-attention mechanism must identify very subtle deviations from a trusted history.

\begin{table}[h!]
	\centering
	\small
	\begin{tabular}{|c|c|c|c|c|c|c|}
		\hline
		\textbf{Threshold ($\tau$)} & \textbf{TP} & \textbf{FP} & \textbf{Recall (\%)} & \textbf{Precision (\%)} & \textbf{F1-Score} & \textbf{Filter Load (\%)} \\ \hline
		$\ge$ 0.65 & 53 & 407 & 93.0 & 11.5 & 0.204 & 31.2 \\ \hline
		$\ge$ 0.70 & 49 & 376 & 86.0 & 11.5 & 0.203 & 28.8 \\ \hline
		$\ge$ 0.75 & 33 & 80  & 57.9 & 29.2 & 0.388 & 7.6 \\ \hline
		$\ge$ 0.80 & 28 & 73  & 49.1 & 27.7 & 0.354 & 6.9 \\ \hline
	\end{tabular}
	\caption{GNN Filtering Performance Across Thresholds}
	\label{tab:gnn_metrics}
\end{table}

\begin{table}[h!]
	\centering
	\small
	\begin{tabular}{|c|c|c|c|}
		\hline
		\textbf{Campaign} & \textbf{Recall (\%)} & \textbf{Precision (\%)} & \textbf{F1-Score} \\ \hline
		C1 & 83.3 & 100 & 90.9 \\ \hline
		C2 & 100 & 100 & 100 \\ \hline
		C3 & 90.0 & 95.0 & 92.4 \\ \hline
		C4 & 100 & 100 & 100 \\ \hline
		C5 & 75.0 & 70.8 & 72.8 \\ \hline
		\textbf{Overall} & \textbf{89.7} & \textbf{92.2} & \textbf{90.9} \\ \hline
	\end{tabular}
	\caption{Co-Attention BERT Performance on $\ge$ 0.70 $\tau$ After GNN Filtering}
	\label{tab:bert_metrics}
\end{table}
\subsection{Per Component Contribution}
As shown in Table \ref{tab:ablation_study}, no single phase is sufficient for robust anomaly detection. Phase 2 (Relational Anomaly) serves as the primary driver of structural recall (61.362\%), confirming that anomalous sender–receiver interaction patterns are the strongest indicator of potential social engineering. However, Phase 2 alone suffers from extremely low precision (2.266\%). The integration of Phase 1 (Volume) and Phase 3 (Community) acts as a necessary regulatory mechanism, filtering out structural noise and raising the combined structural recall to 86.000\%. This demonstrates that our composite structural score successfully isolates a high-risk candidate pool for the subsequent content-based verification stage.
\begin{table}[H]
\centering
\caption{Ablation Study: Structural Phase Performance per component}
\label{tab:ablation_study}
\begin{tabular}{@{}lcccc@{}}
\toprule
\textbf{Configuration} & \textbf{Recall (\%)} & \textbf{Precision (\%)} & \textbf{F1-Score (\%)} \\ \midrule
Phase 1 Only           & 19.767               & 13.427                  & 15.991                 \\
Phase 2 Only           & 61.362               & 2.266                   & 4.371                  \\
Phase 3 Only           & 33.822               & 14.200                  & 20.002                 \\ \midrule
Phase 1 + 2            & 70.614               & 5.549                   & 10.289                 \\
Phase 1 + 3            & 33.822               & 14.200                  & 20.002                 \\
Phase 2 + 3            & 34.505               & 13.919                  & 19.836                 \\ \midrule
\textbf{Phase 1+2+3}   & \textbf{86.000}      & \textbf{11.500}         & \textbf{20.287}        \\ \bottomrule
\end{tabular}
\end{table}
\subsection{Comparison with Existing Methods and Baselines}

To evaluate the necessity of the initial structural filtering stage, we tested a standalone BERT-Only baseline. In this configuration, the co-attention ModernBERT model processed all emails on attack days (7,452 legitimate and 57 attack emails, totaling 7,509 interactions) without prior GNN filtering. Analyzing the entire traffic volume, the standard BERT model achieved a recall of only 63.0\%. While the language model successfully identifies explicit social engineering keywords, it struggles with sophisticated attacks that use benign or contextually normal language. Moreover, applying a deep transformer model to all network traffic introduces an impractical computational overhead. Applying the GNN as a first stage provides critical structural context, improving detection precision while mitigating the high computational cost of analyzing all individual emails.

We also compared our approach against the locality scan statistic ($\Psi$) proposed by Priebe et al. \cite{priebe2005}, which detects anomalies by calculating the Z-score of edge counts within a node's $k=1$ neighborhood using a rolling window. While this method successfully detects the initial onset of novel external attacks, its effectiveness diminishes rapidly due to continuous baseline updating. For instance, in Campaigns 1 and 2, the Z-score peaks significantly on the first day of the attack ($\Psi = 10.00$) but decays steadily on subsequent days (falling to $\Psi = 2.00$ by day 5) as the rolling window incorporates the new edges into normal behavior. Consequently, the scan statistic frequently fails to detect multi-stage grooming tactics and follow-up communications.

This limitation also negatively impacts the detection of low-and-slow campaigns and compromised insider accounts. In Campaign 4 (low-and-slow), the extended temporal gaps between emails caused the scan statistic's rolling window to repeatedly reset, resulting in the model treating a continuous year-long grooming campaign as a series of isolated events rather than a coordinated attack. Similarly, in Campaign 5 (insider threat), the scan statistic registered a significant initial peak ($\Psi = 30.00$) when the compromised user first deviated from historical patterns. However, by the second day of the attack, the rolling window had assimilated this new behavior, causing the anomaly score to drop to $\Psi = 1.30$.

Our proposed framework mitigates these limitations. By integrating the contrast-aware relational score (Phase 2) and linguistic drift metrics ($D_{ling}$) with ModernBERT's co-attention mechanism, the system retains structural and contextual memory throughout the entire duration of the attack.

\begin{table}[htbp]
    \centering
    \small
    \caption{Limitations of Locality Scan Statistic ($\Psi$) across Attack Profiles}
    \label{tab:scan_stat_limits}
    \begin{tabular}{@{}lcc@{}}
        \toprule
        \textbf{Campaign Profile} & \textbf{Initial Peak $\Psi$} & \textbf{Post-Peak $\Psi$} \\ \midrule
        C1 \& C2 (External Burst) & 10.00 (Day 1) & 2.00 (Day 5) \\
        C4 (Low and Slow) & 10.00 (Repeated) & 3.00 -- 4.36 \\
        C5 (Insider Threat) & 30.00 (Day 458) & 1.30 (Day 459) \\ \bottomrule
    \end{tabular}
\end{table}

\subsection{Detection of Insider Threats (Campaign 5)}

Because compromised internal accounts maintain legitimate historical communication patterns, applying the standard global detection threshold ($\tau = 0.70$) resulted in comparatively lower initial performance for Campaign 5, achieving 75.0\% recall and 70.8\% precision. To address this, we implemented the specialized Insider Threat Module, which shifts from a global anomaly detection approach to an ego-centric baseline evaluation. The ITM effectively penalized interactions that deviated from a user's established historical behavior, improving the Campaign 5 detection rate to 83.3\% recall and 78.4\% precision at an adjusted threshold of 0.60. Although further lowering the threshold could theoretically yield 100\% recall, doing so would introduce a prohibitive rate of false positives. The current configuration thus represents an optimal trade-off, demonstrating that integrating linguistic drift ($D_{ling}$) and manipulation intent ($I_{man}$) into a personalized baseline is essential for identifying compromised accounts that evade purely structural filters.

\subsection{Temporal Generalization Analysis}

To evaluate the temporal robustness of our framework, we partitioned the Enron dataset chronologically, training the GraphSAGE model exclusively on data from the 1999–2000 period. The frozen model was then evaluated on unseen data from the 2001–2002 period. To maintain experimental consistency, the synthetic attack campaigns originally injected into the training period were mapped into the 2001–2002 test set. We shifted the campaign timestamps by 731 days, ensuring that the relative duration and email frequency of the attacks remained identical while the background network topology and legitimate traffic patterns evolved. 

The framework demonstrated strong generalization capabilities. At a detection threshold of $\tau = 0.6$, the system achieved a temporal recall of 89.0\%. Even at a more conservative threshold of $\tau = 0.7$, the model maintained a recall of 76.3\%. These results confirm that structural signatures of social engineering such as anomalous relationship formation and communication bursts are fundamentally time-invariant. This validates the use of historical datasets as viable alternative for structural filtering.

\begin{table}[h!]
    \centering
    \small
    \caption{Temporal Generalization Performance (Train: 1999--2000, Test: 2001--2002)}
    \label{tab:temporal_results}
    \begin{tabular}{|c|c|c|c|}
        \hline
        \textbf{Threshold ($\tau$)} & \textbf{Total Attacks} & \textbf{TP} & \textbf{Recall (\%)} \\ \hline
        $\ge 0.60$ & 38 & 34 & 89.0 \\ \hline
        $\ge 0.70$ & 38 & 29 & 76.3 \\ \hline
    \end{tabular}
\end{table}

\subsection{Impact of GNN Filtering and BERT Refinement}
The framework uses a two-stage architecture, where the GNN acts as a high-recall filter to reduce the set of candidate interactions, which is then evaluated by the Co-Attention BERT model for precise detection. Table \ref{tab:gnn_bert_comparison} summarizes the effect of this design on workload, recall, false positives, and precision.

\begin{table}[h!]
	\centering
	\small
	\caption{Impact of GNN Filtering and Co-Attention BERT on Detection Metrics}
\begin{tabular}{|l|p{4.5cm}|p{4.5cm}|}
	\hline
	\textbf{Metric} & \textbf{Stage 1: GNN Only (Filter)} & \textbf{Stage 2: GNN + BERT (Final)} \\ \hline
	Input Volume & 100\% (All Traffic) & 28.8\% (Filtered Candidates) \\ \hline
	Recall (Safety) & 86.0\% & 82.4\% \\ \hline
	False Positives & 376 (High Noise) & 7 (Low Noise) \\ \hline
	Precision & 11.5\% (Weak) & 92.2\% (Strong) \\ \hline
\end{tabular}
\label{tab:gnn_bert_comparison}
\end{table}

The GNN alone achieves high recall (86.0\%) but with significant false positives. After BERT refinement, the majority of false positives are eliminated, and precision reaches 92.2\%. While the Stage 2 recall is slightly lower due to the complexity of the Campaign 5 insider threat, the two-stage design maintains detection reliability while reducing computational load by over 71\%.

\subsection{Summary and Trade-Off Analysis}

Our experiments show a clear trade-off between detecting attacks and avoiding false alarms. The GNN stage catches most suspicious interactions, detecting 86\% of attacks at a 0.70 threshold. However, it also flags many normal emails, like announcements or administrative messages, which lowers precision.

Adding the BERT content analysis stage helps solve this problem. By examining the actual email content and past message context, it removes almost all false positives and raises precision to over 92\%. Some attacks, such as those from insider accounts, remain harder to detect, so recall drops slightly, but overall the system reliably finds attacks.

The two-stage design is also efficient. Only about 29\% of emails need full BERT processing, which saves computation and makes the system practical for real-world use. Overall, separating structural detection and content verification allows the framework to balance safety (high recall), accuracy (high precision), and efficiency in a realistic enterprise environment.

	%--------------------------------------------------
	% References
	%--------------------------------------------------
	\bibliographystyle{plain}
	\bibliography{references}
	
\end{document}